\documentclass[traditabstract]{aa} 
\usepackage{graphicx}
\usepackage{txfonts}
\usepackage{natbib}
\bibpunct{(}{)}{;}{a}{}{,} 

\newcommand{\kms}{km s$^{-1}$}

\begin{document}
   \title{ALMA and VLA observations of recombination lines and continuum 
toward the Becklin-Neugebauer object in Orion}


   \author{Roberto Galv\'an-Madrid\inst{1}
         \and
          Ciriaco Goddi\inst{1}
          \and
          Luis F. Rodr\'iguez\inst{2,3}
          }

\institute{
  European Southern Observatory, Karl-Schwarzschild-Str. 2, 85748 Garching, Germany
  \email{rgalvan@eso.org}
  \and 
  Centro de Radioastronom\'ia y Astrof\'isica, Universidad Nacional Aut\'onoma de M\'exico, Morelia 58090, 
     Mexico
  \and
  Astronomy Department, Faculty of Science, King Abdulaziz University,
P.O. Box 80203, Jeddah 21589, Saudi Arabia
          }

   \date{Received  ; accepted }

 
  \abstract
  {
  Compared to their centimeter-wavelength counterparts, millimeter recombination lines (RLs) 
  are intrinsically brighter and are free of pressure broadening. 
  We report observations of RLs (H$30\alpha$ at 231.9 GHz, H$53\alpha$ at 
42.9 GHz) and the millimeter and centimeter 
continuum toward the Becklin-Neugebauer (BN) object in Orion, obtained 
from the Atacama Large Millimeter/submillimeter Array (ALMA) Science Verification archive 
and the Very Large Array (VLA). The RL emission appears to be arising from the slowly-moving, dense 
($N_e=8.4\times10^6$ cm$^{-3}$) base of the ionized envelope around BN. This ionized gas has a relatively 
low electron temperature ($T_e<4900$ K) and small ($<<10$ km s$^{-1}$) bulk motions. 
Comparing our continuum measurements with previous (non)detections, 
 it is possible that BN has large flux variations in the millimeter. However, dedicated 
observations with a uniform setup are needed to confirm this.
From the H$30\alpha$ line, the central line-of-sight LSR velocity of BN is 26.3 km s$^{-1}$. 
}
  \authorrunning{Galv\'an-Madrid et al.}
  \titlerunning{ALMA and VLA Observations of BN}
  \maketitle
%

\section{Introduction}

Understanding the nature of the radio sources in the Orion
BN/KL region is important because these sources exhibit
significant diverging proper motions that may trace the recent ($\sim 500$ yr ago) 
disintegration of a compact cluster of young massive stars \citep{rodriguez05,gomez08,goddi11b}. 
This past interaction may also have caused the explosive outflow seen in vibrationally-excited 
H$_2$ and CO lines \citep{zapata09,bally11,nissen12}. 

In particular, observing hydrogen recombination lines (RLs)
can be used to obtain the line-of-sight velocities of the radio sources, 
which are needed to complement the proper-motion information. 
Millimeter (mm) RLs are in principle better suited to giving the central velocity 
of the system because they are more optically thin, much less pressure broadened, 
and intrinsically brighter than centimeter (cm) lines \citep[e.g.,][]{gs02,peters12}. 
However, longer wavelength lines are a valuable complement because they are  sensitive to the gas 
density. By observing multiple RL transitions, hints to the structure of the ionized nebula 
can be obtained even in unresolved observations \citep{kzk08,gm09}. 

In this paper we report our findings in the recently released 
Science Verification (SV) data taken toward Orion BN/KL with the Atacama Large Millimeter/submillimeter Array (ALMA). 
We also report on previous RL observations obtained with the Very Large Array (VLA), as well as on 
the continuum obtained with both interferometers.


\section{Data}


Orion BN/KL was observed on January 19, 2012 
with ALMA as part of the SV program.
In this paper we make use of the fully calibrated visibilities 
released on April 12, 2012. The data consisted of 20 different spectral 
windows covering the frequency range from 213.72 GHz to 246.62 GHz. 
Each spectral window had a uniform spectral resolution 
of 0.488 MHz ($\sim0.6$ \kms ) and was observed for 
an on-source time of $\sim 15$ minutes. 
The array used 16 12-m antennas in a compact configuration with baselines in the 
range of 15 k$\lambda$ to 269 k$\lambda$. The pointing center was 
$\alpha_\mathrm{J2000}=05^\mathrm{h}~35^\mathrm{m}~14\rlap.{^\mathrm{s}}35$, 
$\delta_\mathrm{J2000}=-05^\circ~22'~35\rlap.{''}0$. 

The absolute amplitude scale was derived from observations of Callisto and Mars. 
The phase calibrator was J0607--085. We imaged the phase calibrator and 
measured a flux density of $S=1.40$ Jy. The SMA data
base\footnote{http://sma1.sma.hawaii.edu/callist/callist.html/} reports flux
densities for J0607--085 within ten days of the ALMA observation: 
$S_\mathrm{09Jan2012}=1.25\pm0.06$ Jy, $S_\mathrm{27Jan2012}=1.37\pm0.07$ Jy. 
From this, we estimate that the amplitude scale of the ALMA observations 
is accurate to $\sim 10~\%$. The processing was done in CASA versions 3.3.0 and 3.4.0. 

We also present archival Very Large 
Array\footnote{The National Radio Astronomy Observatory is operated by Associated
Universities, Inc. under cooperative agreement with the National Science
Foundation.} (VLA) data. 
The H$53\alpha$ 
($\nu_0=42.95197$ GHz) B-array observations have been previously presented by 
\cite{rodriguez09b} (project AR635). 

Finally, several epochs of continuum observations at 8.4 GHz
in the A configuration were processed using standard methods in the
AIPS software. In all epochs 1331+305 was used as amplitude calibrator
with an adopted flux density of 5.21 Jy. The epochs, project codes,
phase calibrators, and their bootstrapped flux densities are
given in Table \ref{t1}. 

\begin{table}[t]
\caption{\label{t1} VLA calibrators at 8.4 GHz}
\centering
\begin{tabular}{cccc}
\hline
Epoch  &  Project &  Phase  		  & Bootstrapped flux \\
       &          &  calibrator       & density [Jy] \\
\hline
06 Sep 1991   &  AM335  &  0530+135  &  $1.77\pm0.04$  \\
29 Apr 1994   &  AM442  &  0501--019  &  $2.38\pm0.01$  \\
22 Jul 1995   &  AM494  &  0541--056  &  $1.97\pm0.14$  \\
21 Nov 1996   &  AM543  &  0501--019  &  $1.51\pm0.01$  \\
11 Jan 1997   &  AM543  &  0501--019  &  $1.45\pm0.01$  \\
13 Nov 2000   &  AM668  &  0541--056  &  $1.11\pm0.01$  \\
12 May 2006   &  AR593  &  0541--056  &  $1.39\pm0.01$  \\
\hline
\end{tabular}
\end{table}

\section{Results}

Continuum subtraction was done in the u-v domain by carefully selecting the line-free channels 
in each spectral window and fitting a linear spectral baseline.  
Continuum images from different spectral windows are consistent with each other, 
indicating a low contamination from unaccounted line emission. 

All the spectral windows were imaged with a threefold purpose: to detect the 
H$30\alpha$ RL in the BN/KL radio sources (I, BN, and n), 
to detect molecular emission from BN, and to make a continuum image. 
Whereas we succeeded in detecting the RL in BN, it was not detected 
elsewhere. Also, no molecular emission was detected from BN. 
We present maps that have not been corrected for primary beam attenuation. The corrected 
fluxes are quoted in the text and tables.

\subsection{Continuum.}

\begin{figure}[h]
   \centering
   \includegraphics[width=9cm]{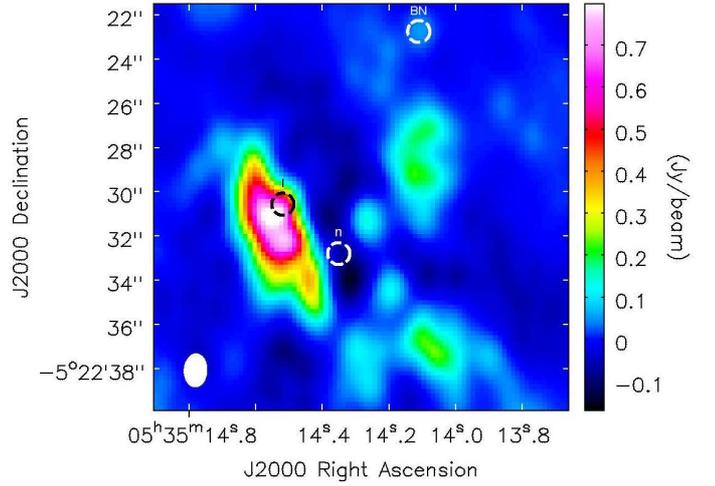}
      \caption{Continuum  ALMA image centered at 230.17 GHz (1.3 mm) of the Orion BN/KL 
region. The centimeter sources I, BN, and n are marked with circles. 
The emission is dominated by the bright, extended emission from the Orion ``Hot Core", 
with a peak intensity of 834 mJy beam$^{-1}$.  
The rms noise is $\sigma \sim$ 10 to 15  mJy. 
The half-power beam width (HPBW) is $1\rlap.{''}49 \times 1\rlap.{''}01$, with 
a position angle PA$=-3.2^\circ$. 
              }
         \label{f1_continuum}
   \end{figure}

The low declination of Orion BN/KL, as well as the bright, extended emission 
dominated by the ``hot core'' (HC), limit   
the dynamic range of synthesis imaging and complicate the detection of weak sources 
in this field at (sub)mm wavelengths. 
Figure \ref{f1_continuum} shows a deep continuum image from the ALMA data obtained 
by co-adding the line-free continuum visibilities from all spectral windows. 
The average frequency (wavelength) in this map is 230.17 GHz (1.30 mm). 
The dynamic range in this image is determined by limitations in the 
u-v coverage, and it is in the range of $\sim 10$ to 15 mJy.  
Our continuum map is $\sim\times 2$ deeper than the released continuum 
image\footnote{http://almascience.eso.org/alma-data/science-verification}, 
and similar in quality to the CARMA 1.3-mm map reported by \citet{fs08}. 

The centimeter radio source I \citep{reid07,goddi11b} is only $\sim 0\rlap.{''}7$ 
from the 1.3-mm peak of the HC, 
which has an extension of $\sim 8\rlap.{''}6 \times 3\rlap.{''}2$ (see Fig. \ref{f1_continuum}).  
Therefore, the continuum emission of source I cannot be resolved from the HC in 
arcsecond-resolution images \citep[e.g.,][]{beuther05,goddi11a}. 
The BN object, as traced by its cm continuum, 
is $\sim 6\rlap.{''}7$ from the hot-core continuum peak. In the map shown in 
Fig. \ref{f1_continuum} we detect BN with a peak intensity of 55 mJy beam$^{-1}$ 
(112 mJy beam$^{-1}$ when primary beam attenuation is taken into account).  
However, it is still at the intensity level of some of the extended emission. 

%
%

\begin{figure}[h]
   \centering
   \includegraphics[width=9cm]{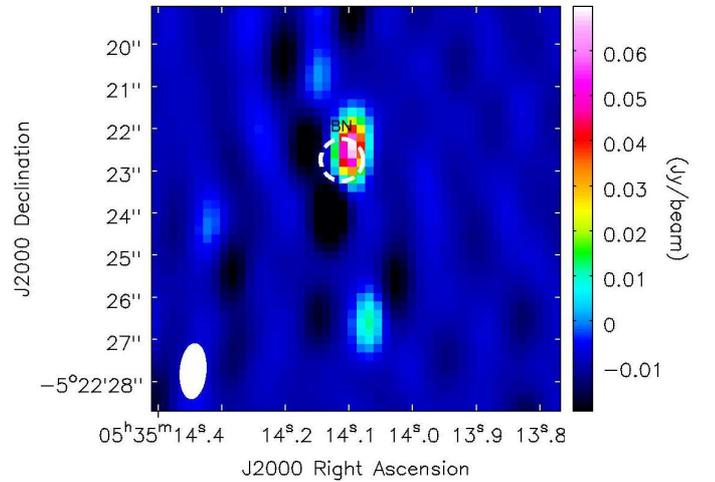}
      \caption{Continuum  ALMA image centered at 231.17 GHz (1.3-mm) of the compact emission 
of the BN object. Only baselines with length $B>100$ k$\lambda$ were used 
to filter out the extended emission in the field. The BN object is detected as an 
unresolved source. The rms noise around BN in this map is 
$\sigma \sim 3$ mJy. HPBW$= 1\rlap.{''}32 \times 0\rlap.{''}62$, PA$=175.5^\circ$. 
             }
         \label{f2_continuum}
   \end{figure}

To disentangle the compact continuum emission from the extended emission in 
the field, we made cleaned maps after removing the short baselines. 
This degradation of the u-v coverage produces artifacts (closely-spaced 
sidelobes) close to source I and the HC, and we refrained from using this 
part of the image. However, at the position of BN, where emission is fainter, 
the restricted u-v coverage effectively removes the extended emission 
around BN.  
We found that a minimum baseline length of $\approx 100$ 
k$\lambda$ gives optimal results, filtering out the emission on scales 
$>1\rlap.{''}4$. Figure \ref{f2_continuum} shows this continuum map around BN. The noise 
is $\times 5$ lower than in Fig. 1 ($\sim 3$ mJy). The 1.3-mm flux density 
of BN 
in this image is 62 mJy. After primary beam correction, the flux density of BN is 
$S_\mathrm{1.3mm}=126$ mJy. 

We note that 
the flux density for BN is the same, within $20~\%$, 
regardless of the u-v coverage restriction applied. 
A faint compact source with 
primary beam corrected flux $S_\mathrm{1.3mm}=26$ mJy 
is also detected $4\rlap.{''}3$
south of BN. This newly detected source is 
 possibly associated with the extended mid-infrared sources IRc6 and IRc6 N
\citep{shuping04}. 

%
%

It is possible that BN is variable in the (sub)mm continuum. While our detection of BN 
is consistent with the 218 GHz flux of $S=150\pm50$ mJy (from OVRO observations in 1993 and 1994)
reported by \citet{plambeck95} and \citet{blake96}, \citet{fs08} do not detect BN at 231 GHz down 
to a $3\sigma$ level of $\sim 34$ mJy (CARMA observations in 2007). 
However, the significance of these variations can be affected by the inhomogeneity of the mm 
observations. 
We have reprocessed three epochs of 
VLA archival data at 3.6 cm (8.4 GHz) to look for possible 
variations in the continuum emission from BN. 
Table \ref{t2} lists the measured 3.6-cm fluxes, together with the four epochs reported by \citet{zapata04},   
also at 3.6 cm, and the available 1-mm band measurements. 
Based on this data set, the 3.6-cm 
flux  is quite stable (within $30~\%$ of the maximum value).

\begin{table}[t]
\caption{\label{t2} BN fluxes.}
\centering
\begin{tabular}{cccc}
\hline
Epoch  &  S$_\mathrm{3.6cm}$ [mJy] &  S$_\mathrm{1.3mm}$ [mJy] & Notes \\
\hline
06 Sep 1991  & $4.61\pm0.20$  & -- & a \\
Nov 1993 to Feb 1994 & -- & $150\pm50$ & b \\
29 Apr 1994  & $3.73\pm0.04$  & -- & c \\
22 Jul 1995  & $3.36\pm0.3$   & -- & c \\
21 Nov 1996  & $3.43\pm0.1$   & -- & c \\
11 Jan 1997  & $3.90\pm0.07$  & -- & c \\
13 Nov 2000  & $3.83\pm0.14$  & -- & a \\
12 May 2006  & $4.78\pm0.11$  & -- & a \\
Mar 2007     & -- & $<34$ & d \\
19 Jan 2012  & -- & $126\pm14$ & e \\
\hline
\end{tabular}
\tablefoot{
a: VLA-A archival data, HPBW$\approx0\rlap.{''}3$ (this paper). 
b: OVRO observations at 218 GHz, HPBW$=1\rlap.{''}5 \times 1\rlap.{''}0$ 
\citep{plambeck95,blake96}. 
c: VLA-A, HPBW$\approx0\rlap.{''}3$ 
\citep{zapata04}. 
d: CARMA observations at 231 GHz, HPBW$=2\rlap.{''}5 \times 0\rlap.{''}8$ 
\citep{fs08}. 
e: ALMA SV data, HPBW$=1\rlap.{''}3 \times 0\rlap.{''}6$ (this paper). The error 
is determined by variations in the obtained flux with different apertures in the 
$B_\mathrm{min}=100$ k$\lambda$ map. 
The flux density has been corrected for primary beam attenuation. 
}
\end{table}

\subsection{Recombination lines}

We report the first detection of mm RL emission toward the BN 
object. We also searched for 
H$30\alpha$ emission in the rest of the 
Orion BN/KL field, in particular from sources I and n, without success. Just as 
with the continuum, the proximity of source I to the HC contaminates any 
emission from the relatively faint RL with brighter, more extended 
molecular lines with rest frequencies within a few MHz of the  H$30\alpha$  line. 
Figure 3 shows the H30$\alpha$ and the H53$\alpha$ 
\citep[data originally presented in ][]{rodriguez09b} 
spectra toward BN. 
The line parameters are summarized in Table \ref{t3}. 
Figure 4 shows the velocity-integrated intensity maps (moment 0) for both lines.
Similar to the mm continuum,  
the H30$\alpha$ map uses only baselines with length $> 100$ k$\lambda$\footnote
{The H30$\alpha$ line is also detected in maps that use the complete baseline range, 
but bright molecular emission from the HC centered at $\sim 10$ MHz blueshifted 
with respect to the RL creates sidelobes that are for the most part removed  
by taking only the longest baselines.}.  
 
 %
   \begin{figure}
   \centering
   \includegraphics[width=9cm]{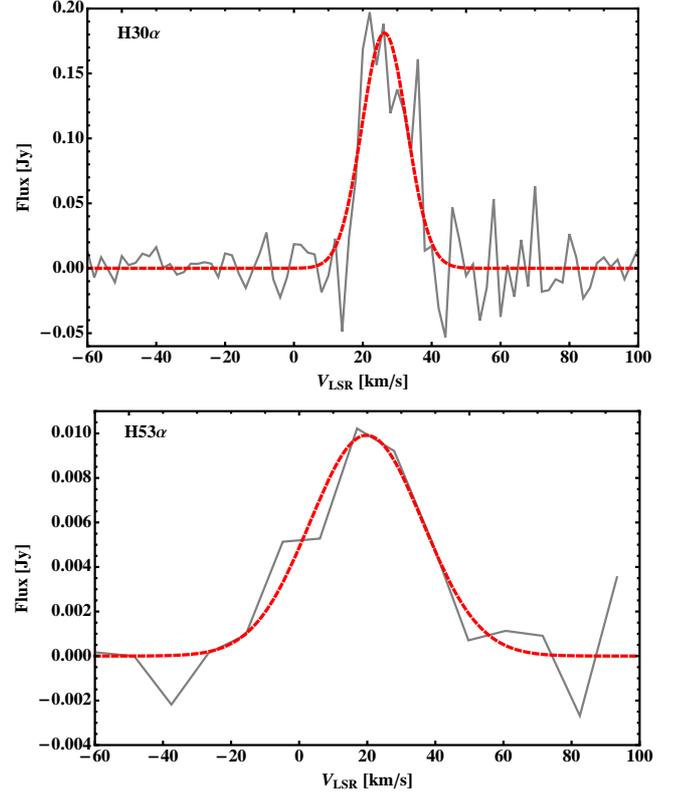}
      \caption{
        Hydrogen recombination line spectra (gray) and Gaussian fits (red) toward BN. 
        The \textit{top} panel shows the H30$\alpha$ line observed with ALMA. 
        The flux is corrected by primary beam attenuation. The \textit{bottom} 
panel shows the H53$\alpha$ line observed with the VLA. The channel widths are 
2.0 km s$^{-1}$ and 10.9 km s$^{-1}$, respectively. 
}
         \label{f3_rrls}
   \end{figure}
%

\begin{table}[h]
\caption{\label{t3} Recombination line parameters.}
\centering
\begin{tabular}{ccccc}
\hline
Transition  &  $\nu_\mathrm{rest}$  & $S_\mathrm{peak}$ &  $v_\mathrm{LSR,peak}$ & FWHM  \\
                  & [GHz]    &  [mJy]  & [km s$^{-1}$]    & [km s$^{-1}$]  \\
\hline
H$30\alpha$  &  231.9009 &  $181\pm14$ & $26.3\pm0.5$ & $15.4\pm1.3$ \\
H$53\alpha$  &  42.95197 &  $9.9\pm1.1$ & $19.8\pm2.1$ & $39.0\pm5.0$ \\
\hline
\end{tabular}
\tablefoot{
From Gaussian fits to the spectra shown in Fig. 3 using an aperture adjusted to the emitting 
area shown in Fig. 4. 
}
\end{table}

 The H$30\alpha$ line is narrower than the H$53\alpha$ line, and it is redshifted by a 
 few km s$^{-1}$. This behavior has been observed in hypercompact HII regions 
 around embedded young massive stars \citep{kzk08,gm09}.  
 Although BN is not deeply embedded in molecular gas, the same interpretation seems to apply: 
 that the difference in linewidth is caused by the action of ``pressure'' broadening 
 from particle collisions in the low-frequency line, whereas the velocity differences are 
 caused by optical depth effects. In an outflowing gas where the denser part is closer to the 
 central star, the optically thinner, higher frequency line will trace the centermost gas 
 from both the front and the back sides of the nebula. In contrast, the optically thicker, 
 lower frequency line, will trace gas farther away from the central star, mainly from the front (blueshifted) 
 side of the nebula \citep[see][]{kzk08}. Therefore, the H$30\alpha$ line should give a better 
 estimation of the line-of-sight velocity of the star inside the BN nebula, 
 $v_\mathrm{LSR}\mathrm{(BN)}=26.3\pm0.5$ km s$^{-1}$.

%
   \begin{figure}
   \centering
   \includegraphics[width=9cm]{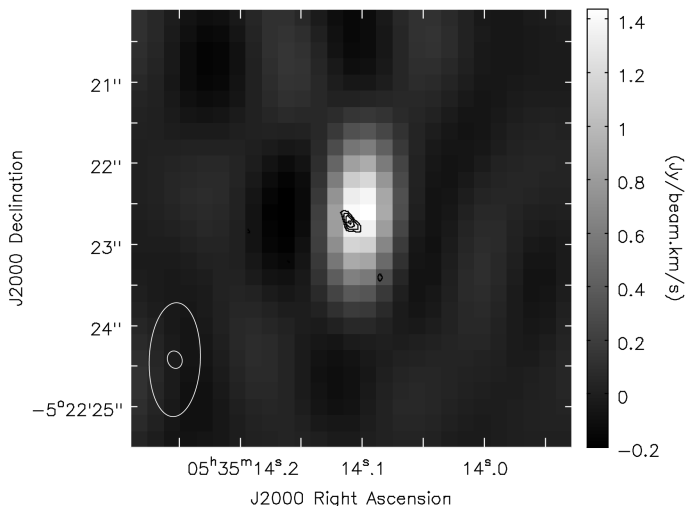}
      \caption{
        Maps of the velocity-integrated (moment 0) emission of the recombination 
lines. The color scale shows the H30$\alpha$ line observed with ALMA 
(HPBW$= 1\rlap.{''}40 \times 0\rlap.{''}63$, PA$=177.9^\circ$). Contours show the 
H53$\alpha$ line observed with the VLA 
(HPBW$= 0\rlap.{''}22 \times 0\rlap.{''}18$, PA$=19.4^\circ$). 
             }
         \label{f4_mom0}
   \end{figure}


The total RL  width ($\Delta v$) has contributions from thermal broadening ($\Delta v_\mathrm{th}$), 
dynamical broadening 
from macroscopic gas motions  ($\Delta v_\mathrm{dy}$), and pressure
broadening ($\Delta v_\mathrm{pr}$). We elaborate this aspect in Appendix A. 
The line FWHM of the combined profile is given by 
\citep[equation 2 of ][]{gm09}: 

\begin{equation}
\Delta v \approx  0.534\Delta v_\mathrm{pr} +(\Delta v_\mathrm{dy}^2 + 
\Delta v_\mathrm{th}^2 + 0.217\Delta v_\mathrm{pr}^2)^{1/2}. 
\end{equation}

\noindent

The H30$\alpha$ line is only broadened thermally and by dynamical motions 
($\Delta v_\mathrm{pr}<0.6$ km s$^{-1}$ from equation A.4). Also, in the BN object this line 
appears to be too narrow (15.4 km s$^{-1}$) even for pure thermal broadening at  $T_e=10^4$ K 
 (equation A.2). 
Therefore, for the gas traced by the H30$\alpha$ line, $T_e$ is at most 4900 K, 
and the bulk gas motions are constrained to be much smaller than the thermal width. 
Finally, using equation 3 of \cite{kzk08}, the average electron density traced by these 
RLs is $N_e=8.4\times10^6$ cm$^{-3}$. 

Non-LTE effects and dust opacity may play a role for (sub)mm RLs. In Appendix B we show that the RLs 
in BN are reasonably close to LTE and that it is possible that dust in the line-of-sight 
makes a considerable contribution to the 1.3-mm continuum.

\section{Discussion}
 
One possible explanation for the differences in the mm fluxes that have been reported for 
BN is that they are, at least partially, caused by systematic errors in the measurements. 
All the compiled cm fluxes are 
measured with the VLA A-array, which has an absolute flux-scale uncertainty of a few percent, 
 while the flux scale of mm interferometers is known to be only accurate to within 
$\sim 10~\%$\footnote{http://www.vla.nrao.edu/astro/calib/manual/, 
http://haneul.astro.illinois.edu/$\sim$wkwon/CARMA/fluxcal/}.  
Differences in u-v coverage may also mimic flux variations, but since the continuum emission 
from BN is less than one arcsecond in size, this effect should not be 
important. Indeed, the flux that we recover from BN 
is the same within $\sim 20~\%$ for different u-v restrictions, (see Section 3.1). 
Alternatively, if BN is variable in the mm, it could be caused by the sudden 
recombination of dense gas \citep[the timescale of which is $\sim$ 1 month for densities 
$N_e \sim 10^6$ cm $^{-3}$,][]{osterbrock89}. 
We emphasize that  
only dedicated multiepoch observations with a uniform setup can clarify this issue. 
 
\smallskip
 
The BN object is known to be a young star with a mass $M_\star \approx 10~M_\odot$ \citep{rodriguez05}. 
In the absence of other forces (e.g., wind acceleration mechanisms), 
this mass is enough to gravitationally confine ionized hydrogen within a radius of $\approx 55$ AU 
\citep[see][]{keto07}. If the H30$\alpha$ is emitted within a volume as small as the radio continuum 
\citep[radius $\sim 25$ AU,][]{rodriguez09b}, this could explain the lack of bulk motions in this 
RL. The H53$\alpha$ line appears to be tracing gas that is outflowing toward 
us with a velocity $\sim 6.5$ km s$^{-1}$ blueshifted with respect to the systemic.  
Infrared and optical RLs have also been detected toward BN and 
extend to velocities $>100$ km s$^{-1}$ \citep{scoville83}. However, these RLs are significantly 
affected by obscuration and non-LTE effects, so they may trace gas from a different volume 
than the radio and (sub)mm RLs.

\section{Conclusions}

We report the first detection of 1.3 mm RL (H30$\alpha$) emission toward Orion from the BN object. 
The mm RL appears to be arising from the dense ($N_e=8.4\times10^6$ cm$^{-3}$), static 
(bulk motions $<<10$ km s$^{-1}$) base of the ionized nebula around the central massive star in BN, 
whose line-of-sight LSR velocity is now 
estimated to be $v_\mathrm{LSR}\mathrm{(BN)}=26.3\pm0.5$ km s$^{-1}$. Compared to the 
H$30\alpha$ line, the optically thicker H$53\alpha$ seems to trace gas outflowing toward 
us at $\sim 6.5$ km s$^{-1}$. 

Dedicated ALMA observations at subarcsecond resolution 
would be able to detect RL emission from the rest of the Orion BN/KL sources, to obtain 
their 3D velocities by combining them with the available proper motion measurements, 
and to quantify their possible flux variations.

\begin{acknowledgements}
R.G.-M. and C.G. acknowledge funding from the European Community's Seventh 
Framework Programme (/FP7/2007-2013/) under grant agreement No. 229517R.
L.F.R. acknowledges the support of DGAPA, UNAM, and CONACyT (Mexico).
This paper makes use of the following ALMA data: ADS/JAO.ALMA\#2011.0.00009.SV. 
ALMA is a partnership of ESO (representing its member states), NSF (USA) and NINS (Japan), 
together with NRC (Canada) and NSC and ASIAA (Taiwan), in cooperation with the Republic of Chile. 
The Joint ALMA Observatory is operated by ESO, AUI/NRAO and NAOJ. 
The authors are grateful to the ALMA staff for doing the observations and calibrating the 
data. The referee, Dr. T. L. Wilson, and the editor, Dr. M. Walmsley, provided prompt and useful 
comments to this manuscript. The authors also thank Dr. Richard Plambeck for providing 
corrections. 
\end{acknowledgements}

\bibliographystyle{aa} 
\bibliography{references} 

\begin{thebibliography}{26}
\expandafter\ifx\csname natexlab\endcsname\relax\def\natexlab#1{#1}\fi

\bibitem[{{Bally} {et~al.}(2011){Bally}, {Cunningham}, {Moeckel}, {Burton},
  {Smith}, {Frank}, \& {Nordlund}}]{bally11}
{Bally}, J., {Cunningham}, N.~J., {Moeckel}, N., {et~al.} 2011, \apj, 727, 113

\bibitem[{{Beuther} {et~al.}(2005){Beuther}, {Zhang}, {Greenhill}, {Reid},
  {Wilner}, {Keto}, {Shinnaga}, {Ho}, {Moran}, {Liu}, \& {Chang}}]{beuther05}
{Beuther}, H., {Zhang}, Q., {Greenhill}, L.~J., {et~al.} 2005, \apj, 632, 355

\bibitem[{{Blake} {et~al.}(1996){Blake}, {Mundy}, {Carlstrom}, {Padin},
  {Scott}, {Scoville}, \& {Woody}}]{blake96}
{Blake}, G.~A., {Mundy}, L.~G., {Carlstrom}, J.~E., {et~al.} 1996, \apjl, 472,
  L49

\bibitem[{{Friedel} \& {Snyder}(2008)}]{fs08}
{Friedel}, D.~N. \& {Snyder}, L.~E. 2008, \apj, 672, 962

\bibitem[{{Galv{\'a}n-Madrid} {et~al.}(2009){Galv{\'a}n-Madrid}, {Keto},
  {Zhang}, {Kurtz}, {Rodr{\'{\i}}guez}, \& {Ho}}]{gm09}
{Galv{\'a}n-Madrid}, R., {Keto}, E., {Zhang}, Q., {et~al.} 2009, \apj, 706,
  1036

\bibitem[{{Goddi} {et~al.}(2011{\natexlab{a}}){Goddi}, {Greenhill},
  {Humphreys}, {Chandler}, \& {Matthews}}]{goddi11a}
{Goddi}, C., {Greenhill}, L.~J., {Humphreys}, E.~M.~L., {Chandler}, C.~J., \&
  {Matthews}, L.~D. 2011{\natexlab{a}}, \apjl, 739, L13

\bibitem[{{Goddi} {et~al.}(2011{\natexlab{b}}){Goddi}, {Humphreys},
  {Greenhill}, {Chandler}, \& {Matthews}}]{goddi11b}
{Goddi}, C., {Humphreys}, E.~M.~L., {Greenhill}, L.~J., {Chandler}, C.~J., \&
  {Matthews}, L.~D. 2011{\natexlab{b}}, \apj, 728, 15

\bibitem[{{G{\'o}mez} {et~al.}(2008){G{\'o}mez}, {Rodr{\'{\i}}guez}, {Loinard},
  {Lizano}, {Allen}, {Poveda}, \& {Menten}}]{gomez08}
{G{\'o}mez}, L., {Rodr{\'{\i}}guez}, L.~F., {Loinard}, L., {et~al.} 2008, \apj,
  685, 333

\bibitem[{{Gordon} \& {Sorochenko}(2002)}]{gs02}
{Gordon}, M.~A. \& {Sorochenko}, R.~L., eds. 2002, Astrophysics and Space
  Science Library, Vol. 282, {Radio Recombination Lines. Their Physics and
  Astronomical Applications}

\bibitem[{{Jim{\'e}nez-Serra} {et~al.}(2011){Jim{\'e}nez-Serra},
  {Mart{\'{\i}}n-Pintado}, {B{\'a}ez-Rubio}, {Patel}, \& {Thum}}]{js11}
{Jim{\'e}nez-Serra}, I., {Mart{\'{\i}}n-Pintado}, J., {B{\'a}ez-Rubio}, A.,
  {Patel}, N., \& {Thum}, C. 2011, \apjl, 732, L27

\bibitem[{{Keto}(2007)}]{keto07}
{Keto}, E. 2007, \apj, 666, 976

\bibitem[{{Keto} {et~al.}(2008){Keto}, {Zhang}, \& {Kurtz}}]{kzk08}
{Keto}, E., {Zhang}, Q., \& {Kurtz}, S. 2008, \apj, 672, 423

\bibitem[{{Nissen} {et~al.}(2012){Nissen}, {Cunningham}, {Gustafsson}, {Bally},
  {Lemaire}, {Favre}, \& {Field}}]{nissen12}
{Nissen}, H.~D., {Cunningham}, N.~J., {Gustafsson}, M., {et~al.} 2012, \aap,
  540, A119

\bibitem[{{Osterbrock}(1989)}]{osterbrock89}
{Osterbrock}, D.~E. 1989, {Astrophysics of gaseous nebulae and active galactic
  nuclei}

\bibitem[{{Peters} {et~al.}(2012){Peters}, {Longmore}, \&
  {Dullemond}}]{peters12}
{Peters}, T., {Longmore}, S.~N., \& {Dullemond}, C.~P. 2012, \mnras, 425, 2352

\bibitem[{{Plambeck} {et~al.}(1995){Plambeck}, {Wright}, {Mundy}, \&
  {Looney}}]{plambeck95}
{Plambeck}, R.~L., {Wright}, M.~C.~H., {Mundy}, L.~G., \& {Looney}, L.~W. 1995,
  \apjl, 455, L189

\bibitem[{{Reid} {et~al.}(2007){Reid}, {Menten}, {Greenhill}, \&
  {Chandler}}]{reid07}
{Reid}, M.~J., {Menten}, K.~M., {Greenhill}, L.~J., \& {Chandler}, C.~J. 2007,
  \apj, 664, 950

\bibitem[{{Rodr{\'{\i}}guez} {et~al.}(2005){Rodr{\'{\i}}guez}, {Poveda},
  {Lizano}, \& {Allen}}]{rodriguez05}
{Rodr{\'{\i}}guez}, L.~F., {Poveda}, A., {Lizano}, S., \& {Allen}, C. 2005,
  \apjl, 627, L65

\bibitem[{{Rodr{\'{\i}}guez} {et~al.}(2009){Rodr{\'{\i}}guez}, {Zapata}, \&
  {Ho}}]{rodriguez09b}
{Rodr{\'{\i}}guez}, L.~F., {Zapata}, L.~A., \& {Ho}, P.~T.~P. 2009, \apj, 692,
  162

\bibitem[{{Scoville} {et~al.}(1983){Scoville}, {Kleinmann}, {Hall}, \&
  {Ridgway}}]{scoville83}
{Scoville}, N., {Kleinmann}, S.~G., {Hall}, D.~N.~B., \& {Ridgway}, S.~T. 1983,
  \apj, 275, 201

\bibitem[{{Shuping} {et~al.}(2004){Shuping}, {Morris}, \& {Bally}}]{shuping04}
{Shuping}, R.~Y., {Morris}, M., \& {Bally}, J. 2004, \aj, 128, 363

\bibitem[{{Smirnov} {et~al.}(1984){Smirnov}, {Sorochenko}, \&
  {Pankonin}}]{smirnov84}
{Smirnov}, G.~T., {Sorochenko}, R.~L., \& {Pankonin}, V. 1984, \aap, 135, 116

\bibitem[{{Walmsley}(1990)}]{walmsley90}
{Walmsley}, C.~M. 1990, \aaps, 82, 201

\bibitem[{{Wilson} {et~al.}(2009){Wilson}, {Rohlfs}, \&
  {H{\"u}ttemeister}}]{toolsofradio}
{Wilson}, T.~L., {Rohlfs}, K., \& {H{\"u}ttemeister}, S. 2009, {Tools of Radio
  Astronomy} (Springer-Verlag)

\bibitem[{{Zapata} {et~al.}(2004){Zapata}, {Rodr{\'{\i}}guez}, {Kurtz}, \&
  {O'Dell}}]{zapata04}
{Zapata}, L.~A., {Rodr{\'{\i}}guez}, L.~F., {Kurtz}, S.~E., \& {O'Dell}, C.~R.
  2004, \aj, 127, 2252

\bibitem[{{Zapata} {et~al.}(2009){Zapata}, {Schmid-Burgk}, {Ho},
  {Rodr{\'{\i}}guez}, \& {Menten}}]{zapata09}
{Zapata}, L.~A., {Schmid-Burgk}, J., {Ho}, P.~T.~P., {Rodr{\'{\i}}guez}, L.~F.,
  \& {Menten}, K.~M. 2009, \apjl, 704, L45

\end{thebibliography}

\Online

\begin{appendix} 

\section{Linewidth of recombination lines}
RLs are broadened by several mechanisms: a natural broadening  
from the quantum uncertainty $\Delta E$ of the energy level $E$, a 
thermal/microturbulence Gaussian broadening from the motions of the 
emitting particles and of parcels of gas that are much smaller than the beam, 
the Stark (or ``pressure'') broadening from the perturbation of the atomic 
energy levels by the electric field of neighboring charged particles, and a
dynamical broadening from bulk flows (e.g., infall, rotation, outflow) in 
the gas.  
A detailed discussion of these processes and the physics of RLs can be found 
in \cite{gs02}.

\smallskip

For $n > 20$ 
(where H$n\alpha$ is a $n+1 \rightarrow n$ transition), 
the natural linewidth in velocity units is 

\begin{equation}
\Delta v_\mathrm{N} \approx 1.2 \times 10^{-6} c \frac{\ln (n+1)}{(n+1)^2}, 
\end{equation}

\noindent
where $c$ is the speed of light. For H30$\alpha$ and H53$\alpha$, $\Delta v_\mathrm{N}$ 
is $1.3 \times 10^{-3}$ km s$^{-1}$ and $0.5 \times 10^{-3}$ km s$^{-1}$, respectively. 
The natural broadening is negligible for radio and (sub)mm RLs. 

\smallskip

The thermal distributions of velocities of  both the individual particles and small pockets 
of gas (``microturbulence'') produce a Gaussian contribution to the broadening. 
Neglecting microturbulence, the thermal FWHM is  

\begin{equation}
\Delta v_\mathrm{th} = \biggl ( 8 \ln 2 \frac{k_\mathrm{B} T_e}{m_\mathrm{H}} \biggr )^{1/2},  
\end{equation}

\noindent
where $k_\mathrm{B}$ is the Boltzmann constant, $m_\mathrm{H}$ is the hydrogen-atom mass, 
and it is assumed that all the gas is thermalized to the electron temperature $T_e$. 
For a pure-hydrogen gas with $T_e=9000$ K, $\Delta v_\mathrm{th}=20.3$ km s$^{-1}$. 

\smallskip 

The pressure broadening has a Lorentzian shape, and it increases with density and 
quantum number. 
Collisions with ions and electrons 
make different contributions. The contribution from ions has the form
\citep{gs02}  

\begin{equation}
\Delta v_\mathrm{pr,i} \approx \biggl ( N_i \frac{c}{\nu_0}  \biggr ) 
(0.06 + 2.5\times10^{-4}T_e) \biggl ( \frac{n+1}{100} \biggr )^{\gamma_i} 
\biggl ( 1 + \frac{2.8 \Delta n}{n+1} \biggr ), 
\end{equation}

\noindent
where $\gamma_i = 6-2.7\times10^{-5} T_e -0.13(n+1)/100$. For $T_e=9000$ K 
and $N_i=10^7$ cm$^{-3}$, the H$30\alpha$ line is virtually free from ion 
broadening (0.04 km s$^{-1}$, and decreases close to linearly with decreasing 
$T_e$), whereas the H$53\alpha$ line is broadened by 5.1 km s$^{-1}$. 

The collisions with electrons dominate the collisions with ions under most conditions. 
They produce a broadening of width \citep{smirnov84}: 

\begin{equation}
\Delta v_\mathrm{pr,e} \approx \biggl ( 8.2N_e \frac{c}{\nu_0}  \biggr ) 
 \biggl ( \frac{n+1}{100} \biggr )^{4.5} 
\biggl ( 1 + \frac{2.25 \Delta n}{n+1} \biggr ).  
\end{equation}

\noindent
For $N_e=10^7$ cm$^{-3}$, the electron broadenings for the H53$\alpha$ and 
H$30\alpha$ are $\approx 37.3$ km s$^{-1}$  and 0.6 km s$^{-1}$, respectively.  

\smallskip

Finally, the last source of broadening is bulk motions ($\Delta v_\mathrm{dy}$) of the ionized gas, 
which 
may be in the form of outflows/winds, infall/accretion, and/or rotation. 
For unresolved observations, the method outlined here permits estimating  
the magnitude of these motions from the nonthermal linewidth of the line 
that is free of pressure broadening. Of course, ideally one wishes to 
spatially resolve these motions. 
ALMA in its Cycle 1 is now able to resolve the ionized-gas motions in sources as faint as $\sim 10$ 
mJy at the line peak at subarcsecond resolution.

Including all the contributions to the broadening, the 
total linewidth will have a Voigt profile with FWHM given by 
equation (1) in the text. 

\end{appendix}

\begin{appendix} 

\section{cm and (sub)mm recombination lines}
Here we discuss the advantages and limitations that cm and (sub)mm RLs have compared to each other. 
The cm lines are intrinsically fainter and optically thicker, and can be partially absorbed 
by the relatively high continuum opacity at their wavelengths 
\citep[see e.g.,][]{toolsofradio}.  
A possible drawback of (sub)mm RLs is that the free-free 
continuum may be contaminated by dust emission. Another problem is that they are out of
LTE more easily than cm lines, so their interpretation may require careful modeling 
\citep[e.g.,][]{js11,peters12}. 
As shown below, our simple LTE interpretation seems to be reasonable, but it is possible 
that a significant fraction of the 1.3-mm continuum comes from dust in the line of sight. 

\smallskip

Let $b_n$ be the ratio of the population of level $n$ to its LTE population. Then the actual 
line absorption coefficient $\kappa_\mathrm{L}$ is related to the LTE coefficient 
$\kappa_\mathrm{L,LTE}$ by \citep{gs02} 

\begin{equation}
\kappa_\mathrm{L} = \kappa_\mathrm{L,LTE} b_n \beta_n, 
\end{equation}

\noindent
with

\begin{equation}
\beta_n\approx \frac{b_{n+1}}{b_n} \biggl ( 1-\frac{k_\mathrm{B}T_e}{h\nu} \frac{d \ln b_{n+1}}{dn} 
\Delta n \biggr). 
\end{equation}

\noindent
While $0<b_n<1$ expresses the non-LTE level depopulation, $\beta_n$ can be negative when 
there are conditions for maser amplification. 

In the optically-thin case ($\tau_\mathrm{C,1.3mm}\sim0.08$ from the data), the non-LTE 
corrected line intensity $I_\mathrm{L}$ is \citep{gs02} 

\begin{equation}
I_\mathrm{L} \approx I_\mathrm{L,LTE}b_{n+1}(1-\tau_\mathrm{C}\beta/2). 
\end{equation}

Using the tabulated values in the calculations of \cite{walmsley90},  $b_{30}\approx 0.98$  
and $\beta_{30} \sim -1.5$ for a density $N_e=10^7$ cm$^{-3}$.  Therefore, the intensity 
of the H$30\alpha$ line is only amplified by $\sim 3~\%$. 
The H$53\alpha$ is even closer to LTE, with  $b_{53}\approx 0.999$ and $\beta_{53} \sim 0.5$.

\smallskip

Sub(mm) RLs are more opticaly thin and intrinsically brighter than 
cm RLs. 
For a given $T_e$ and $N_e$, and in the Rayleigh-Jeans regime ($h\nu << k_\mathrm{B} T_e$), 
the line absorption coefficient decreases linearly with frequency 
$\kappa_\mathrm{L} \propto \nu^{-1}$, so the LTE line emission coefficient 
$j_\mathrm{L}=\kappa_\mathrm{L}B_\nu(T_e) \propto \nu$. Similarly, the continuum absorption 
coefficient $\kappa_\mathrm{C} \propto \nu^{-2.1}$, so the LTE continuum emissivity 
$j_\mathrm{C} \propto \nu^{-0.1}$. Therefore, in the optically thin regime, the 
line-to-continuum ratio increases almost linearly with 
frequency $S_\mathrm{L} \Delta v /S_\mathrm{C} \propto \nu^{1.1}$.  
Under these assumptions, the expected ratio of the H30$\alpha$ to H$53\alpha$ line-to-continuum 
ratios is 6.4. However, the observed value is 1.6. 
If we assume that only $\sim 32$ mJy of the 1.3-mm 
continuum are due to free-free then the line-to-continuum ratio would increase to the expected value. 
This suggests that dust contributes a significant fraction to the (sub)mm continuum of BN. 
However, it is known (and we have confirmed this with the ALMA data set) that BN is not embedded 
in dense molecular gas. Therefore, it is unlikely that this line-of-sight dust arises in a 
dense core around BN. 
\end{appendix}

\end{document}